\documentclass[sigconf]{acmart}

\usepackage{color-edits}
\usepackage{hyperref}
\usepackage{bbm}
\usepackage{amsmath}

\DeclareMathOperator*{\argmax}{arg\,max}

\addauthor{dh}{orange}

\AtBeginDocument{%
  \providecommand\BibTeX{{%
    \normalfont B\kern-0.5em{\scshape i\kern-0.25em b}\kern-0.8em\TeX}}}

\setcopyright{acmcopyright}
\copyrightyear{2021}
\acmYear{2021}

\acmDOI{}

\acmISBN{}

\acmConference[KDD MIS2 '21]{MIS2 workshop at KDD 2021}{August 14-18th, 2021}{Virtual}

\begin{document}

\title{Recurrent Graph Neural Networks for Rumor Detection in Online Forums}

\author{Di Huang}
\authornote{Work completed during Google Research summer internship.}
\email{dh_599@usc.edu}
\affiliation{%
  \institution{University of Southern California}
  \city{Los Angeles}
  \state{California}
  \country{USA}
}
\author{Jacob Bartel}
\email{bartel@google.com}
\affiliation{%
  \institution{Google}
  \city{Mountain View}
  \state{California}
  \country{USA}
}
\author{John Palowitch}
\email{palowitch@google.com}
\affiliation{%
  \institution{Google Research}
  \city{San Francisco}
  \state{California}
  \country{USA}
}

\renewcommand{\shortauthors}{Huang et al.}

\begin{abstract}
The widespread adoption of online social networks in daily life has created a pressing need for effectively classifying user-generated content. This work presents techniques for classifying linked content spread on forum websites -- specifically, links to news articles or blogs -- using user interaction signals alone. Importantly, online forums such as Reddit do not have a user-generated social graph, which is assumed in social network behavioral-based classification settings. Using Reddit as a case-study, we show how to obtain a derived social graph, and use this graph, Reddit post sequences, and comment trees as inputs to a Recurrent Graph Neural Network (R-GNN) encoder. We train the R-GNN on news link categorization and rumor detection, showing superior results to recent baselines. Our code is made publicly available at \url{https://github.com/google-research/social_cascades}.
\end{abstract}





\maketitle

\section{Introduction}

As online social media becomes increasingly present in peoples' daily lives, greater proportions of users get news and journalistic content directly from their "feed" on accounts like Facebook, Twitter, YouTube, and Reddit. Pew research reported that social media outpaced print news as a news source in the United States in 2018 \cite{shearer2018social}. Some social media platforms are also news-centric -- for instance, a 2016 Pew study found that seven out of ten Reddit users use the platform to get their news \cite{barthel2016seven}. Following these findings, the seminal work \cite{vosoughi2018spread} showed that user interaction signals -- differential patterns of liking, re-sharing, and commenting -- distinguish between posts that link to certain categories of online content, in particular content later identified as "rumors" versus other content.

Due to these phenomena, a recent sub-field of applied machine-learning research has grown, focusing on graph-based artificial intelligence models for classifying links and content shared on social media, particularly for rumor detection \cite{bondielli2019survey}.
Most rumor detection models recently introduced have been tuned for and evaluated on data from online \emph{social networks} like Twitter or Facebook \cite[e.g.][]{wu2018tracing,rosenfeld2020kernel}. These platforms have a natural social graph created by users, which provides an inherent graph on which a Graph Neural Network can propagate rumor information. However, relatively less attention in rumor detection research has been given to online \emph{forums} like Reddit. Our work addresses two nuances specific to forums. First, most forums do not have a natural who-follows-who social graph. Second, most forums do not feature a "repost" option on their platform, preventing usual inter-user cascades seen in social networks \cite{vosoughi2018spread}. Instead, each article in a forum is posted a limited number of times, each time independently by users across the platform. This means that each forum post (unlike social network posts) is the start of its own discussion cascade consisting of a long comment-tree graph, as opposed to repost/share behavior found on social networks.

To address these nuances, we provide a two-fold contribution to this space. First, we illustrate the construction of \emph{emergent} social networks from online forum data, which we use both for feature learning and the downstream neural network computational graph via a GNN. Second, we introduce a Recurrent-GNN model which can well-handle the independent, sequential nature of article posting on forums. Our approach combines an RNN, to capture time-order of posts, with a Graph Neural Network (GNN), to capture the post comment relations of users. We evaluate our approach on the publicly-available Reddit corpus, testing two classification tasks: topic classification and rumor detection. We address preliminaries in Section 2, detail our methods in Section \ref{sec:method}, describe our evaluation experiments in Section \ref{sec:experiments}, and conclude with a brief discussion in Section 5.

\begin{figure*}[!htb]
  \centering
  \includegraphics[width=0.96\linewidth]{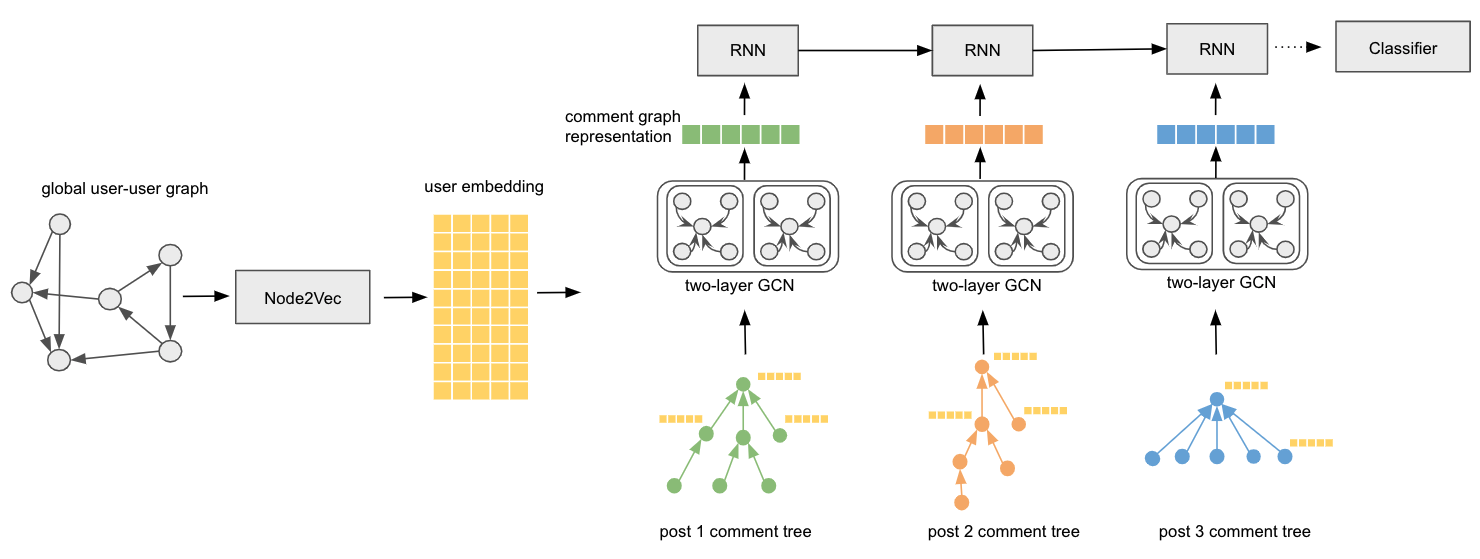}
  \caption{Recurrent Graph Neural Network(R-GNN) Framework. User features are derived from Node2Vec \citep{grover2016node2vec} graph embeddings. For a given article, post representations are learned from commenter features by a GCN component, which are fed to RNN cells. The RNN encodes the article representation its post representations, modelling the article's temporal propagation.}
  \label{gcnrnn}
\end{figure*}

\section{Preliminaries}
News and opinion pieces are being produced with record-breaking volume, and links to these pieces propagate swiftly on social sites like Twitter, Facebook, and Reddit. Formally, a link $m$ is propagated by a sequence of posts $\mathbf{p} = \{p_{1}, p_{2}, \ldots\}$ having corresponding authors $\mathbf{a} = \{a_{1}, a_{2}, \ldots\}$. On forum sites like Reddit, for which our main contributions are intended, each post also has a sequence of \emph{commenters}\, $\mathbf{c} = \{c_{1}, c_{2}, \ldots\}$. Note that $\mathbf{a}, \mathbf{c}\subseteq \mathbbm{U}$ where $\mathbb{U}$ is the total user set of the forum. For each task, we pair each message $m$ with a categorical label $y$, and train various models to predict $y$ given feature data associated with any $\mathbf{p}$, $\mathbf{a}$, and $\mathbf{c}$.

Classification tasks having the above general setting, in which a label $y$ is inferred from a some social media information piece $m$, have attracted great attention from researchers in recent times. For instance, \cite{ott2013negative} uses sentiment analysis to distinguish spam comments and trustworthy comments, and \cite{shu2017fake} detect fake news via assessment on credibility of article headlines. However, content-based classification in this domain can be challenging, especially in the context of rumor detection, as pieces with differing labels can nonetheless feature similar topics and writing styles. \citep{zhou2019fake}. Because of this, other approaches like Traceminer \citep{wu2018tracing} and CSI \citep{ruchansky2017csi} have been proposed which learn from user interaction signals and information propagation paths. Similarly, graph embedding methods such as Node2Vec \citep{grover2016node2vec}, SDNE \citep{wang2016structural} and graph convolutional networks -- GCNs  -- have been widely used for network analysis and graph feature extraction in social network studies. In this work, we propose a framework which combines GCNs (Graph Convolutional Networks \cite[e.g.][]{kipf2016semi,hamilton2017inductive,velivckovic2017graph}) and RNNs (Recurrent Neural Networks) to model the information diffusion process of article links.

We note that the pipeline and model we introduce for our Reddit case-study, as with many of the approaches listed above, can be used for any supervised classification tasks featuring labelled posts or article links. When applied to a task like rumor detection, it is crucial to note that a machine-learning model cannot be used to establish or predict an ultimate verdict on whether a piece of content is true, false, or of high/low quality. As detailed in Section 4, we derive a rumor label for an article by its presence on a fact-check site, regardless of verdict. In this case, our hypothesis is simply that user interaction signals in social media data can be correlated, via deep learning, to articles' potentials to be controversial in the specific sense that they are noticed by a fact-check site. Since we train our model (and baseline models) on fact-check site data, our results are subject to any biases or errors in that data. Nevertheless, as seen in our experimental results, behavioral signals are useful in this regard, and thus our illustration of methods in this space may interest other researchers studying these phenomena.

\section{Method}\label{sec:method}
In this work we introduce a Recurrent Graph Neural Network (R-GNN), illustrated in Figure \ref{gcnrnn}, which models \emph{two} information diffusion processes on forums. First, a graph convolutional network (GCN) encodes features from commenters under each post. Second, a recurrent neural network (RNN) learns a link representation from the sequence of post encodings. We detail these components and feature construction in the next sections.

\subsection{Global User-User Interaction Graph}
Unlike social network (SN) platforms like Facebook and Twitter, forums like Reddit commonly do not have a natural user-friendship graph. To replace the natural graph used in SN studies \citep[e.g.][]{wu2018tracing,bian2020rumor}, we \emph{derive} a graph $\mathbbm{G}$ from user interactions. Formally, $\mathbbm{G} = (\mathbbm{U}, \mathbbm{E})$ where $\mathbbm{U}$ is the user set and $\mathbbm{E}$ is the edge set. $\mathbbm{E}$ is a set of undirected, weighted edges $(\{u_i, u_j\}, w_{ij})$ where $w_{ij}$ is the count of comment-replies or post-replies between $u_i$ and $u_j$, on \emph{any} post from Reddit (including those not contained in our link dataset). This graph represents \emph{proximal} friendships between users given their commenting activity. We encode these friendships as feature inputs to our model by computing user graph embeddings on $\mathbbm{G}$ with node2vec \cite{grover2016node2vec}.

\subsection{GCN Post Encoding}
In addition to the global graph $\mathbbm{G}$, we also construct a \emph{local reply-graph} $G_p$ for each post $p$. The graph $G_p = (U_p, E_p)$ consists of the users $U_p$ who commented on $p$ (including the author), and each weighted edge represents the number of times each pair of commenters replied to each other. With the aforementioned graph embeddings as user features, we encode each post $p$ into a hidden vector $\mathbf{v}_p$ with a two-layer GCN \cite{kipf2016semi} applied to $G_p$.

\subsection{RNN+GCN Post-Sequence Encoding}
We formulate inference on a link $m$ as a temporal sequence classification problem on its time-ordered posts $\{p_{m1}, p_{m2}, \ldots\}$. At each timestep, we encode the post with the comment-graph GCN, and pass that representation to an RNN unit. Finally, we predict $y$ with a multi-layer perceptron (MLP) applied to the RNN encoding, as illustrated in Figure 1. Formally, given a link $m$ and its corresponding post sequence $p_1, p_2, \ldots$, we apply the GCN to obtain post encodings $\mathbf{v}_1, \mathbf{v}_2, \ldots$, and infer a predicted $\hat{y}$ as
\begin{equation}
    \hat{y} = \argmax \text{MLP}(\text{RNN}(\mathbf{v}_1, \mathbf{v}_2, \ldots))
\end{equation}

\section{Experiments}\label{sec:experiments}
In this section we describe the evaluation of our R-GNN model against five baselines on two tasks: article categorization and rumor detection.
Four of our baselines are standard (non-neural) machine learning methods applied to simplified features. Our fifth baseline is an established RNN-based method called TraceMiner, which has been previously evaluated on similar tasks using Twitter data \citep{wu2018tracing}. We evaluate two versions of our R-GNN against these baselines. First, we remove the GCN component from our model, concatenating all authors $\mathbf{a}$ and commenter sequences $\{\mathbf{c}_1, \mathbf{c}_2, \ldots\}$ associated with a post into a single sequence, which we feed to an RNN. We refer to this version as ``R-GNN(-replygraph)", as it removes the influence of the comment graph signal from the learning process. This provides an ablation study of R-GNN to better evaluate the combination of GCN and RNN components in our proposed approach. Second, we evaluate the full R-GNN as described in Section 3. Finally, we note that all of our experiments implicitly test our hypothesis that a ``proximal" friendship graph can be derived from user interactions as a useful signal in these tasks, as described in Section 3. All-told, we design our experiments to answer three main research questions as follows:
\begin{itemize}
    \item[\textbf{RQ1}] Can signals derived purely from user interactions (absent a natural social graph) be successful in classifying links that are shared in online forums?
    \item[\textbf{RQ2}] Can diffusion process modeling with deep neural networks outperform standard ML models, applied to online forums?
    \item[\textbf{RQ3}] Can our RNN+GCN hybrid model outperform simpler RNN-only baselines, especially for rumor detection?
\end{itemize}


\subsection{Baseline Methods}
Here we describe baseline models against which we compare RGNN. The hyperparameters of all models, including both variants of RGNN, were tuned on a 10\% validation set and tested on a 10\% test set. The tables in this section report test set metrics.

\textbf{SVM/XGBoost}. To address RQ1, we compare R-GNN and TraceMiner with SVM and XGboost. We apply SVM and XGBoost directly to the average embedding vectors of all users that authored or commented on any post with a given link $m$. This provides a "shallow" model baseline against the neural models.

\noindent\textbf{Traceminer \cite{wu2018tracing}}. Traceminer is a RNN-based diffusion model. It directly uses the \emph{post-author} graph embedding as the post representation for  RNN input. Importantly, Traceminer does not use any commenter or comment-tree information. We label this model using "Traceminer(author)".   Furthermore, R-GNN(-replygraph) can show whether commenters' information can improve the performance compared with Traceminer baseline, which only contains authors.

\subsection{Link Categorization}
For the link categorization task, we match links from the UCI News Aggregator Dataset~\footnote{https://archive.ics.uci.edu/ml/datasets/News+Aggregatorused}~\cite{Dua:2019} to Reddit posts which embed those links. Explicitly, for each link $m$ in the UCI News Aggregator data, $y$ is the topic label, and $\mathbf{p}$ is the sequence of Reddit posts which embed $m$. News links in this data are divided into four news categories: business, science/technology, entertainment and health. Based on the selected 8,220 URLs and their associated posts, we construct a global user network with 77.2k nodes and 153.6k edges.

In Table \ref{categorization}, the best performance is notated in bold and the second best score is underlined. We see that on the categorization task, SVM with author and commenter embedding ranks first based on micro-F1 score while XGoost with author and commenter embedding achieves the best macro-F1 score. However, both R-GNN(-replygraph) and Traceminer(author) are outperformed by standard baselines, and our full R-GNN achieves only comparable results. Among the diffusion-based deep learning methods, the performance of R-GNN(-replygraph) and R-GNN both surpass the baseline Traceminer(author). In addition, models with both authors' and commenters' embedding have the better results than models with solely authors information. Thus, we can see that commenters' features are highly useful for categorizing types of news.


\begin{table}[!htb]\small
\centering
\caption{URL Categorization.}
\begin{tabular}{c|c|c}
\hline
Model & micro-F1 & macro-F1 \\ 
\hline
SVM(author) & 0.5757 & 0.4548 \\
XGBoost(author) & 0.5697 & 0.4649 \\
SVM(author+commenter) & \textbf{0.5953} & \underline{0.4770} \\
XGoost(author+commenter) & \underline{0.5903} & \textbf{0.4834}\\
Traceminer(author) & 0.5182 & 0.4055\\
\hline
R-GNN(-replygraph) & 0.5487 & 0.4423 \\
R-GNN & 0.5243 & 0.4768 \\
\hline
\end{tabular}
\label{categorization}
\end{table}

\subsection{Rumor Detection}
Fact-checking websites aggregate evidence for or against particular claims made by news articles and (sometimes) social media posts. Whether or not it is true, such content can be interpreted as a rumor since it caused doubt and suspicion during its propagation process. In our rumor detection task setup, we regard all the news links which show up on Snopes~\footnote{https://www.snopes.com/}, Politifact~\footnote{https://www.politifact.com/} and Emergent~\footnote{http://www.emergent.info/} as rumor news. The complete dataset of links is maintained at  Kaggle~\footnote{https://www.kaggle.com/arminehn/rumor-citation}. To find non-rumors, we use negative sampling to extract the same amount of news links from the UCI dataset to produce the URL categorization task. All told, we build a dataset of 7,352 news links, with an equal amount of postive "rumor" examples and negative examples. The global network built from the authors and commenters on Reddit posts containing these links has 201.1k nodes and 413.0k edges.

Table \ref{rumor} shows that our R-GNN model has the highest F1 score and our R-GNN(-replygraph) achieves the highest accuracy. SVM(author) and Traceminer(author) ranks the second on accuracy and F1 score respectively. Overall, sequential modeling with deep learning achieved better performance than non-neural baselines on this task.



\begin{table}[!htb]\small
\centering
\caption{Rumor Detection.}
\begin{tabular}{c|c|c}
\hline
Model & Accuracy & F1 \\ 
\hline
SVM(author) & 0.6963 & 0.7025 \\
XGBoost(author) & \underline{0.6908} & 0.6886 \\
SVM(author+commenter) & 0.6790 & 0.6447 \\
XGoost(author+commenter) & 0.6646 & 0.6594 \\
Traceminer(author) & 0.6401 & \underline{0.7536} \\
\hline
R-GNN(-replygraph) & \textbf{0.7057} & 0.7485\\
R-GNN & 0.6609 & \textbf{0.7731}  \\
\hline
\end{tabular}
\label{rumor}
\end{table}

\subsection{Analysis}
Returning to our three research questions, we note that the features for all models were computed solely from a user-interaction based \emph{derived} proximal friendship graph, described in Section 3. As all models performed far better than random chance on each task, we can answer RQ1 in the affirmative, that this derived graph provides a useful signal for link classification in online forums.

On the link categorization task, interestingly, non-neural baselines outperformed TraceMiner and R-GNN. For this experiment, we can tentatively answer RQ2 in the negative. This could be because the diffusion/user-reply processes for the standard news links in UCI Aggregator data news URLs across categories may be similar, and thus non-informative to the categorization task. However, an interesting find that arose from this experiment was that methods that included \emph{commenter} features strongly outperformed those that did not (including R-GNN models vs TraceMiner). We conjecture that is likely due to the strong forum-community signals provided by the commenters' node2vec graph embeddings. 

On the other hand, for the rumor detection task, our full R-GNN model outperformed both neural and non-neural baselines, and R-GNN(-replygraph). This suggests that the diffusion/user-reply processes which feed into the GCN are more useful signals for this task. Thus in this case we can answer RQ2 and RQ3 in the affirmative.

\section{Discussion}
In this work we introduced an approach for rumor detection and (more generally) link classification on forum websites, and evaluated this approach on Reddit data. Our model, a Recurrent Graph Neural Network (RGNN), is able to capture both the diffusion process of each link through post comment-graphs via a GCN, and simultaneously the sequential nature of link-posting on forums via an RNN. When applied to a link topic categorization task, our approach had superior performance to other RNN-based methods, but had comparable performance to non-neural baselines. When applied to a rumor detection task, our approach had superior performance to all baselines. To our knowledge, this is the first appearance of an RGNN in this space, and among the first demonstrations of deep learning on online user interactions without a natural social graph.

Automated rumor detection via artificial intelligence (and more generally, online content categorization) in social networks is a growing area of research, featuring a rich landscape of model architectures and classification tasks. In this short paper we have examined a narrow subset of potential tasks, model architectures, and available features in this space. For instance, to better understand the effect of various \emph{interaction}-based graph signals -- which let the model learn from diffusion processes on the directed graph of user actions -- we have disregarded the many \emph{content}-based signals, e.g. text or images, available for the tasks in our paper and others. However, in doing so, we have shed light on the capacity of state-of-the-art GNNs to model article sharing in forums with interaction-based features alone. This exposes headroom to improve GNN architectures and interaction-based feature construction for classification tasks.

\bibliographystyle{ACM-Reference-Format}
\bibliography{main}










\end{document}